\documentclass[aps,11pt,prd,longbibliography,notitlepage,nofootinbib,floatfix]{revtex4-1}
\usepackage{tikz}
\usepackage{feynmp-auto}
\usepackage{amsmath}
\usepackage{todonotes}
\usepackage[utf8]{inputenc}
\usepackage[T1]{fontenc} 
\usepackage{array}
\usepackage{bbm}
\usepackage{subcaption}
\usepackage{soul}	
\usepackage{xcolor}
\usepackage{graphicx}
\usepackage{epsfig}
\usepackage{color}
\usepackage{slashed}
\usepackage{comment}
\usepackage{epstopdf}
\usepackage{xspace}
\usepackage{mathrsfs}
\usepackage[euler]{textgreek}
\usepackage{amsmath}
\usepackage{amsthm}
\usepackage{amsfonts}
\usepackage{amssymb}
\usepackage{graphicx}
\usepackage[font={small}]{caption}
\usepackage{subcaption}
\usepackage{float}
\usepackage{multirow}
\usepackage[hang, flushmargin,bottom]{footmisc} 
\usepackage{placeins}
\usepackage{enumitem}
\usepackage{slashed}
\usepackage{bbm}
\usepackage{appendix}
\usepackage{tabularx}
\usepackage{siunitx}

\usepackage{titlesec}
\titleformat{\chapter}[display]
  {\normalfont\LARGE\bfseries}
  {\chaptertitlename\ \thechapter}{5pt}{\LARGE}
  \titlespacing*{\chapter}{0pt}{-20pt}{35pt}
\usepackage{bigstrut}
\usepackage{pst-node}
\usepackage{pstricks}
\usepackage{physics}
\usepackage{mathtools}
\usepackage{setspace}
\usepackage{fancyhdr}
\usepackage[makeroom]{cancel}
\usepackage{feyn}
\newcommand{\be}{\begin{equation}}
\newcommand{\ee}{\end{equation}}
\newcommand{\bes}{\begin{equation*}}
\newcommand{\ees}{\end{equation*}}

\usepackage{hyperref}
\hypersetup{%
  colorlinks = true,
  linkcolor  = blue,
  citecolor  = blue,
  urlcolor   = blue,
}
\usepackage{soul}
\usepackage{xpatch}
\makeatletter
\xpretocmd{\todo}{\@bsphack}{}{}
\xapptocmd{\todo}{\@esphack}{}{}
\makeatother

\newcommand{\beq}{\begin{equation}}
\newcommand{\eeq}{\end{equation}}

\usepackage{tikz}

\DeclareRobustCommand{\swatch}[1]{\tikz[baseline=-0.6ex]\node[fill=#1,shape=rectangle,draw=black,thick,minimum width=5mm,rounded corners=0.5pt](){};}

\newcommand{\pT}{\ensuremath{p_\mathrm{T}}\xspace}
\newcommand{\cls}{\ensuremath{\text{CL}_s}\xspace}
\newcommand{\MET}{\ensuremath{p_T^\mathrm{miss}}\xspace}

\newcommand{\tanb}{\ensuremath{\tan\beta}\xspace}
\newcommand{\ttbar}{\ensuremath{t\bar{t}}\xspace}
\newcommand{\MH}{\ensuremath{M_{\mathrm{H}}}\xspace}
\newcommand{\Mh}{\ensuremath{M_{\mathrm{h}}}\xspace}
\newcommand{\MA}{\ensuremath{M_{\mathrm{A}}}\xspace}
\newcommand{\MN}{\ensuremath{M_{\mathrm{N}}}\xspace}

\newcommand{\GeV}{\text{GeV}}

\newcommand{\herwig}{H\protect\scalebox{0.8}{ERWIG}\xspace}
\newcommand{\rivet}{R\protect\scalebox{0.8}{IVET}\xspace}

\newcommand{\contur}{\textsc{Contur}\xspace}

\newcommand{\madgraph}{\textsc{MadGraph5}\xspace}

\newcommand{\leff}{\ensuremath{\lambda_{\mathrm{eff}}}\xspace}

\definecolor{green}{HTML}{008000}
\definecolor{goldenrod}{HTML}{DAA520}
\definecolor{magenta}{HTML}{FF00FF}
\definecolor{silver}{HTML}{C0C0C0}
\definecolor{indigo}{HTML}{4B0082}
\definecolor{skyblue}{HTML}{87CEEB}
\definecolor{darkgoldenrod}{HTML}{B8860B}
\definecolor{orange}{HTML}{FFA500}
\definecolor{yellow}{HTML}{FFFF00}
\definecolor{saddlebrown}{HTML}{8B4513}
\definecolor{blue}{HTML}{0000FF}
\definecolor{turquoise}{HTML}{40E0D0}
\definecolor{yellow}{HTML}{FFFF00}
\definecolor{white}{HTML}{FFFFFF}
\definecolor{whitesmoke}{HTML}{F5F5F5}
\definecolor{hotpink}{HTML}{FF69B4}

\newcommand{\myComment}[1]{}

\begin{document}
\title{\Large{Probing Matter Unification through Higgs Physics at the LHC}}
\author{Mark Brettell$^{1,2}$, Jon Butterworth$^{1}$, 	
Hridoy Debnath$^{3}$, Pavel Fileviez P{\'e}rez$^{3}$}
\affiliation{
  $^{1}$Department of Physics and Astronomy, University College London, Gower St., London, WC1E 6BT, UK \\
  $^{2}$Particle Physics Department, Rutherford Appleton Laboratory, Didcot, UK\\
  $^{3}$Physics Department and Center for Education and Research in Cosmology and Astrophysics, Case Western Reserve University, Cleveland, OH 44106, USA}
\email{mark.brettell@cern.ch, j.butterworth@ucl.ac.uk, hxd253@case.edu, pxf112@case.edu}
\date{\today}

\begin{abstract}
  We investigate the Higgs sector phenomenology of the minimal low scale quark--lepton unification theory.
  In addition to the Standard Model-like Higgs boson, the theory predicts a new CP-even neutral Higgs, a
  CP-odd neutral Higgs, and two charged Higgses, together with heavy right-handed neutrinos and leptoquarks.
  We analyze the scalar spectrum, the Yukawa structure responsible for charged-fermion masses, and the neutrino
  sector realized through an inverse-seesaw mechanism, emphasizing that the couplings of the new Higgs bosons are
  directly tied to the structure of matter unification and therefore lead to characteristic decay patterns dominated
  by third-generation fermions.
  We compute the branching ratios and production cross sections of the new Higgs states at the Large Hadron Collider
  (LHC) and assess their impact on present and future collider measurements. We find that current LHC data probe
  significant regions of parameter space for the neutral heavy Higgs bosons, with the strongest sensitivity arising
  from channels involving tau leptons, tops, and related final states, while the charged Higgs remains less
  constrained because of its electroweak production mechanism.
  Our results show that Higgs measurements provide a powerful and complementary probe of low-scale matter unification
  beyond direct leptoquark searches and open a new avenue for testing quark--lepton unification at present and
  future colliders, with discovery potential in high-luminosity LHC running.
\end{abstract}
\maketitle
\section{INTRODUCTION}
The Standard Model (SM) has been extraordinarily successful in describing the known elementary particles and their interactions. Its gauge structure has been tested with high precision in many experiments, and the discovery of the
Higgs boson at the Large Hadron Collider (LHC)~\cite{ATLAS:2012yve,CMS:2012qbp} confirmed the Brout-Englert-Higgs
mechanism~\cite{Englert:1964et,Higgs:1964ia,Higgs:1964pj} of electroweak symmetry breaking.
So far, the measured properties of the Higgs boson are consistent with SM expectations, further strengthening the theory as the correct description of particle physics up to the electroweak scale.

Despite these successes, the SM leaves several fundamental questions unanswered. It does not explain the origin of neutrino masses,
the observed matter--antimatter asymmetry, the nature of dark matter, or the pattern of fermion masses and mixings.
These shortcomings motivate the search for physics beyond the SM at the TeV scale.
At the LHC, this program includes both direct searches for new particles and indirect probes through precision measurements of
previously-unobserved SM processes and of the properties of the Higgs boson.
Since the Higgs sector is directly tied to mass generation, it is a particularly sensitive probe of theories that address the flavor
structure of the SM or unify quarks and leptons.

Quark--lepton unification~\cite{Pati:1974yy,Pati:1973rp} is one of the most appealing ideas for physics beyond the SM. In these theories, quarks and leptons are embedded into common multiplets of a larger gauge symmetry, leading naturally to new states that connect the two sectors. While such unification is often associated with very high scales, it was shown in Ref.~\cite{FileviezPerez:2013zmv} that a minimal realization can occur near the TeV scale. This low-scale theoretical framework predicts a rich collider phenomenology, including scalar leptoquarks, heavy right-handed neutrinos, and an extended Higgs sector. The phenomenology of the leptoquark sector has been studied in detail in Refs.~\cite{FileviezPerez:2022fni,FileviezPerez:2022rbk,FileviezPerez:2021arx,FileviezPerez:2021lkq,Flacke:2025xwl,FileviezPerez:2023rxn,Gedeonova:2022iac,Murgui:2021bdy,Miralles:2019uzg,Faber:2018qon,Butterworth:2025ttw}. In particular, in Ref.~\cite{Butterworth:2025ttw} we showed that current LHC data already place significant constraints on the leptoquark masses, while leaving viable regions of parameter space, especially when leptoquark decays involve the heavy-neutrino sector.
The collider signatures of the scalar and vector leptoquarks are very important to test the idea of quark-lepton unification,
but the minimal theory predicts also a rich scalar sector with interesting properties.

The scalar sector of the minimal theory for matter unification has received less attention. In addition to the SM-like Higgs boson, the theory contains a second Higgs doublet and therefore predicts a heavy CP-even neutral Higgs, a CP-odd neutral Higgs, and two charged Higgs bosons. The couplings of these states are controlled by the same structure responsible for quark--lepton unification, implying characteristic decay patterns and correlations among quark and lepton final states. The heavy-neutrino sector can further modify the phenomenology by opening additional visible or invisible decay channels. As a result, Higgs observables provide an important and complementary probe of low-scale quark--lepton unification. This theory predicts unique relations between the Higgs decays~\cite{FileviezPerez:2021arx}, when we neglect the fermion masses, for example, one 
has that for  $\tanb \ll1$, 
\begin{eqnarray}
    \Gamma(H (A)\to \bar{\tau}\tau) = 3 \Gamma(H (A)\to  \bar{b} b), \nonumber
\end{eqnarray}
while in the opposite regime, when $\tanb \gg 1 $, one obtains the following relation:
\begin{eqnarray}
    \Gamma(H (A)\to \bar{\tau}\tau) = \frac{1}{3} \Gamma(H (A)\to  \bar{b}b). \nonumber
\end{eqnarray}
Here, $H$ and $A$ are the new neutral CP-even and CP-odd Higgses, respectively. The parameter \tanb is defined by the ratio between the vacuum expectation values of the two Higgs doublets present in the theory.
Clearly, these relations can help us to test the idea of quark-lepton unification at colliders.

The main purpose of this paper is to study in detail the Higgs signatures in the context of the minimal low scale quark--lepton unification theory at the LHC. We analyze the scalar spectrum, the couplings and decay modes of the new Higgs bosons, and the impact of the heavy-neutrino sector on Higgs phenomenology. We study the invisible decay of the SM-like Higgs boson and the resulting constraints on the relevant parameters. In order to understand the collider bounds, we then compute the branching ratios and production cross sections of the new CP-even, CP-odd neutral Higgses and charged Higgs states and investigate the regions of parameter space that can be constrained by current and future LHC measurements. We find that di-tau measurements play a crucial role in constraining a large fraction of parameter space when \tanb is very small or very large. 
Our results show that Higgs physics provides a sensitive test of low-scale quark--lepton unification that is complementary to direct leptoquark searches.

This paper is organized as follows. In Sec.~\ref{sec:model} we review the minimal theory for quark-lepton unification, with emphasis on the fermion masses, neutrino sector, and Higgs interactions. In Sec.~\ref{sec:bsmhiggs} we discuss the implications for the decays of the SM-like Higgs boson. In Sec.~\ref{sec:contur} we study the production cross sections for the new scalars, and their impact on current and future collider measurements. We summarize our main conclusions in Sec.~\ref{sec:summary}.
\section{MATTER UNIFICATION}
\label{sec:model}
The idea of quark-lepton unification is one of the best ideas we have for physics beyond the Standard Model.
Typically, the idea of matter unification is realized at a high-scale, far beyond the reach of current or even future colliders.
However, one can define a simple theory for quark-lepton unification at the multi-TeV scale~\cite{FileviezPerez:2013zmv} if neutrino masses are generated via the inverse seesaw mechanism.
This theoretical framework for minimal quark-lepton unification at the low scale is based on the gauge group~\cite{FileviezPerez:2013zmv}
\begin{equation*}
    SU(4)_C \otimes SU(2)_L \otimes U(1)_R,
\end{equation*}
where the SM fermionic fields are unified in the following representations 
\begin{eqnarray}
F_{q_L} &=&
\left(
\begin{array}{cccc}
u_r & u_g & u_b  & \nu 
\\
d_r & d_g & d_b  & e
\end{array}
\right)_L \sim (\mathbf{4}, \mathbf{2}, 0), \\[1ex]
F_{u_R} &=&
\left(
\begin{array}{cccc}
u_r & u_g & u_b & \nu
\end{array}
\right)_R \sim (\mathbf{{4}}, \mathbf{1}, 1/2), 
\\[1ex]
 F_{d_R} &=&
\left(
\begin{array}{cccc}
d_r & d_g & d_b & e
\end{array}
\right)_R \sim (\mathbf{{4}}, \mathbf{1}, -1/2).
\end{eqnarray}
Here $u_i$ and $d_i$ (with $i=r,g,b$) are the SM quarks with different colors. This matter unification occurs for each SM fermionic family.

The Higgs sector is composed of the following fields:
\begin{eqnarray}
     H_1 &=& \begin{pmatrix}
         H_1^+\\
         H_1^0
     \end{pmatrix} \sim (\mathbf{1},\mathbf{2},1/2) , \hspace{1 cm} \Phi = \begin{pmatrix}
         \Phi_8 & \Phi_4\\
         \Phi_3 &0
     \end{pmatrix} + \sqrt{2} \ T_4 \ H_2 \sim (\mathbf{15},\mathbf{2},1/2),  \nonumber 
\end{eqnarray}
and
\begin{eqnarray}
      \chi &=& \begin{pmatrix}
         \chi_c & \chi_0
     \end{pmatrix} \sim (\mathbf{4},\mathbf{1},1/2). 
     \label{Higgs1} \nonumber
 \end{eqnarray}
Here the $\Phi$ field is needed to achieve a realistic relation between charged fermion masses, $H_2\sim (1,2,1/2)$ is the second Higgs doublet, $\Phi_3$ and $\Phi_4$ are the scalar leptoquarks, $\Phi_8$ is a colored octet, and $\chi_c$ is a Goldstone boson ``eaten'' by the vector leptoquark. In the above equation, $T_4={\rm{diag}}(1,1,1-3)/2\sqrt{6}$ is a $SU(4)_C$ generator. 

The vacuum expectation value (VEV) of the scalar field $\chi$, $\left<\chi\right> = (0,0,0,v_\chi/\sqrt{2})$, is responsible for spontaneous breaking of the gauge symmetry: 
\begin{displaymath}
SU(4)_C \otimes SU(2)_L \otimes U(1)_R \Longrightarrow SU(3)_C \otimes SU(2)_L \otimes U(1)_Y
\end{displaymath}
Notice that $v_\chi$ defines the quark-lepton unification scale and mass of the vector leptoquark present in the theory. For more details, see the detailed discussion in Ref.~\cite{Debnath:2026zow}, where we study the lower bound on the symmetry breaking scale from flavour violating processes.

The effective scalar potential for the two Higgs doublets $H_1$ and  $H_2$ is given by 
\begin{eqnarray}
    \mathcal{V}(H_1 ,H_2) &= & m_{11}^2  H_1^\dagger H_1 +m_{22}^2  H_2^\dagger H_2 -\left[ m_{12}^2  H_1^\dagger H_2 + h.c. \right] + \frac{\lambda_1}{2} \left( H_1^\dagger H_1\right)^2 \nonumber 
\end{eqnarray}
\begin{eqnarray}
    & + &\frac{\lambda_2}{2} \left(H_2^\dagger H_2\right)^2 +
     \lambda_3( H_1^\dagger H_1)( H_2^\dagger H_2)+ \lambda_4 ( H_1^\dagger H_2)( H_2^\dagger H_1) \nonumber \\ &+&\left[\frac{\lambda_5}{2}( H_1^\dagger H_2)^2 +\lambda_6 ( H_1^\dagger H_1)( H_1^\dagger H_2) +\lambda_7 ( H_2^\dagger H_2)( H_2^\dagger H_1) + \text{h.c.} \right].
\end{eqnarray}
Notice that this scalar potential is quite general. However, the $SU(4)_C$ symmetry constrains the Higgs couplings to quarks and leptons.
The physical fields for the CP-even Higgs are defined as 
\begin{eqnarray}
   \begin{pmatrix}
       H \\ h 
   \end{pmatrix} = \begin{pmatrix}
       \cos{\alpha}  & \sin{\alpha}\\
       -\sin\alpha & \cos{\alpha}
   \end{pmatrix} \begin{pmatrix}
       H_1^0 \\H_2^0
   \end{pmatrix},
\end{eqnarray}
while the physical fields for the CP-odd Higgses can be written as 
\begin{eqnarray}
    \begin{pmatrix}
       G \\ A 
   \end{pmatrix} = \begin{pmatrix}
       \cos{\beta}  & \sin{\beta}\\
       -\sin\beta & \cos{\beta}
   \end{pmatrix} \begin{pmatrix}
       A_1^0 \\A_2^0
   \end{pmatrix},
\end{eqnarray}
and the physical fields for the charged Higgses are given by 
\begin{eqnarray}
    \begin{pmatrix}
       G^{\pm} \\ H^{\pm} 
   \end{pmatrix} = \begin{pmatrix}
       \cos{\beta}  & \sin{\beta}\\
       -\sin\beta & \cos{\beta}
   \end{pmatrix} \begin{pmatrix}
       H_1^\pm \\H_2^\pm
   \end{pmatrix}.
\end{eqnarray}
In the above equations, $G$ and $G^\pm$ are the Goldstone bosons eaten by the $Z^0$ and $W^\pm$, respectively.
In the decoupling limit, $M_H \gg M_h$, the masses of the physical Higgses can be written as 
\begin{eqnarray}
    M_H^2 &=& \MA^2 + ( \lambda_H+ \lambda_A)   v^2  , \hspace{1 cm}  M_h^2  =  \lambda_h  v^2, \\
    M_A^2 &=& \frac{m_{12}^2}{s_\beta c_\beta} -  \lambda_A v^2 , \ {\rm{and}} \
    M_{H^{\pm}}^2 = M_A^2 + \frac{v^2}{2} (\lambda_5 -\lambda_4  ),
\end{eqnarray}
where $\tanb = v_2/v_1$. Here $v_1$ and $v_2$ are VEVs of the neutral components of the Higgs doublets $H_1$ and $H_2$, respectively.
See Appendix~\ref{appendix} for more details. See also \cite{FileviezPerez:2021arx,FileviezPerez:2022fni} for a detailed discussion of the Higgs sector and the correlation between the different interactions.

The Yukawa couplings for the charged fermions with the Higgses $H_1$ and $H_2$ can be written as 
\begin{eqnarray}
    - \mathcal{L}_{Y} & \supset &  \bar{Q}_L \left(Y_1 \Tilde{H_1} + Y_2 \frac{1}{2\sqrt{3}} \Tilde{H_2}\right)u_R  \ + \bar{\ell}_L \left(Y_1 \Tilde{H_1} - Y_2 \frac{\sqrt{3}}{2} \Tilde{H_2}\right)\nu_R \nonumber \\
    & + & \bar{Q}_L \left(Y_3H_1 + Y_4 \frac{1}{2\sqrt{3}} H_2\right)d_R +\bar{\ell}_L \left(Y_3H_1 - Y_4 \frac{\sqrt{3}}{2} H_2\right)e_R + \text{h.c.},
\end{eqnarray}
 where $\Tilde{H}_j = i \sigma_2 H^*_j$ with $j=1,2$. After the electroweak symmetry breaking, the masses for the SM  charged fermions can be written as 
  \begin{eqnarray}
    M_u &=& Y_1 \frac{ v_1}{\sqrt{2}} + \frac{1}{2 \sqrt{3}} Y_2 \frac{ v_2}{\sqrt{2}}, \\
    M_d &=& Y_3  \frac{v_1}{\sqrt{2}} + \frac{1}{2 \sqrt{3}} Y_4 \frac{ v_2}{\sqrt{2}}, \\
    M_e &=& Y_3\frac{ v_1}{\sqrt{2}} - \frac{\sqrt{3}}{2} Y_4 \frac{ v_2}{\sqrt{2}}.
    \label{eq:fermionmasses}
  \end{eqnarray}
   The mass matrices for the charged fermions can be diagonalized as follows 
  \begin{eqnarray}
      M_U^{\text{diag}} = U_L^\dagger M_U U_R , \hspace{0.5 cm} M_d^{\text{diag}}= D_L^\dagger M_dD_R , \hspace{0.5cm} M_E^{\text{diag}}= E_L^\dagger M_E E_R.
  \end{eqnarray}
See Appendix~\ref{appendix} for the Feynman rules where these matrices appear once we write all interactions in the physical basis. 
Although the minimal model predicts Dirac neutrino masses, achieving the observed tiny neutrino masses would require extreme cancellation or fine-tuning. The Dirac mass term for the neutrinos is given by 
\begin{eqnarray}
    M_\nu^D = Y_1 \frac{ v_1}{\sqrt{2}} - \frac{3}{2 \sqrt{6}} Y_2 \frac{ v_2}{\sqrt{2}}
    \label{eq:Dirac}.
\end{eqnarray}
One can avoid the fine-tuning in the neutrino sector and generate a small Majorana mass for neutrinos by extending the fermionic sector with three left-handed singlet fields, $S_L \sim (1,1,0)$ and implementing the inverse seesaw mechanism~\cite{Mohapatra:1986bd}. The relevant Lagrangian can be written as~\cite{FileviezPerez:2013zmv} 
\begin{eqnarray}
    \mathcal{L}_\nu^D \supset \  Y_5 \bar{F}_{u_R} \chi S_L + \frac{1}{2} \mu S_L^T CS_L  + \text{h.c.}.
\end{eqnarray}
 In the  ($\nu$, $\nu^c$, $S$) basis, the neutrino mass matrix takes the form
  \begin{equation}
    \left( \nu \  \nu^c \  S  \right) 
    \left(\begin{array}{ccc} 
      0 & M_\nu^D & 0  \\ 
      (M_\nu^D)^T & 0 & M_\chi^D \\
      0 &  (M_\chi^D)^T & \mu
    \end{array}\right)  
    \left(\begin{array}{c} \nu \\  \nu^c \\ S  \end{array}\right).
  \end{equation}
  Here $M_\nu^D$ is given by Eq.~\eqref{eq:Dirac} and 
  $
  M_\chi^D = Y_5 \, v_\chi / \sqrt{2}.
  $
  When $M_\chi^D \gg M_\nu^D \gg \mu$ holds, the active neutrino mass is given by 
  \begin{equation}
    m_\nu \approx \mu \, (M_\nu^D)^2 / (M_\chi^D)^2.
  \end{equation}
  Therefore, one can have very small neutrino masses without fine-tuning. Here, $\mu$ is protected by a global symmetry and it can be small to achieve the correct neutrino masses~\cite{FileviezPerez:2013zmv}. A similar model for quark-lepton unification without the neutrino masses mechanism was studied in Refs.~\cite{Smirnov:1995jq,Smirnov:2018ske,Valencia:1994cj}.
\subsection{Right-handed Neutrino Decays}
\begin{figure}[hbt]
     \centering
     \includegraphics[width=0.6\linewidth]{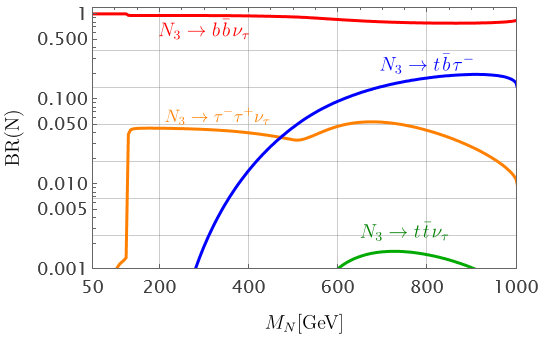}
     \caption{Branching ratios of the decays of 3rd generation right-handed neutrinos as a function of mass.
       Here we assume the same mass for the leptoquarks $\Phi_3$ and $\Phi_4$ at 1~TeV,
       $\MH =\MA = M_{H^+} =500 $ GeV , $y_2=y_4=1, \kappa=\tanb=\leff = 0.1$ and the mixing between right and
       SM neutrinos to be $10^{-4}$.}
     \label{n3br}
\end{figure}
This quark lepton unification framework predicts a rich decay phenomenology for the right-handed neutrinos.
Since in this theory, the active neutrino masses are generated via the inverse seesaw mechanism,
the mixing between the active and heavy neutrinos is given by
\begin{eqnarray}
    V_{\text{mix}} \approx \sqrt{\frac{m_\nu}{\mu }}.
\end{eqnarray}
Therefore, generically the mixing between the active and right-handed neutrinos is small. For instance, when $\mu \sim O(10^{-2})$ GeV, the corresponding mixing angle is of the order of $10^{-4}$. 
The heavy neutrinos in this theory have interesting decays:
\begin{eqnarray}
    N \to \phi_3^{1/3} \bar{b} ,\  \phi_3^{-2/3} u ,\ H^\pm e^\mp_j , \ A  \nu_j,\  H \nu_j, \ h \nu_j, \ Z \nu_j, \ W^\pm e^\mp_j. 
\end{eqnarray}
Here, $\phi_3^{1/3}$ and $\phi_3^{-2/3}$ are scalar leptoquarks~\cite{Butterworth:2025ttw}. 
Notice that in the scenarios where the leptoquarks are heavier than the right-handed neutrinos,
the 3-body decays mediated by the leptoquarks can be very important.
In Fig.~\ref{n3br} we show the numerical predictions for the branching ratios of the decays of third generation
right-handed neutrinos as a function of its mass.
Here we assumed the  mass of the leptoquarks $\Phi_3$ and $\Phi_4$ are $1$ TeV, $\MH =\MA = M_{H^+} =500 $ GeV,
$y_2=y_4=1, \kappa=\tanb=\leff = 0.1$ and mixing between right and SM neutrinos to be $10^{-4}$. For the definition of $y_2$ and $y_4$ see the study in Ref.~\cite{Butterworth:2025ttw}.
Leptoquarks and the Higgses mediate the decays of the right-handed neutrinos, and decays into 3rd-generation quarks
and leptons are dominant. In these benchmark scenarios, one has interesting results, the decays mediated by the
Higgs dominate. Notice that these decays can change the dedicated searches in which one looks for heavy neutrinos at colliders. 

\subsection{New Decays of the SM-like Higgs}
 \begin{figure}[hbt]
    \centering
    \begin{subfigure}[b]{0.48\textwidth}
        \centering
        \includegraphics[width=\linewidth]{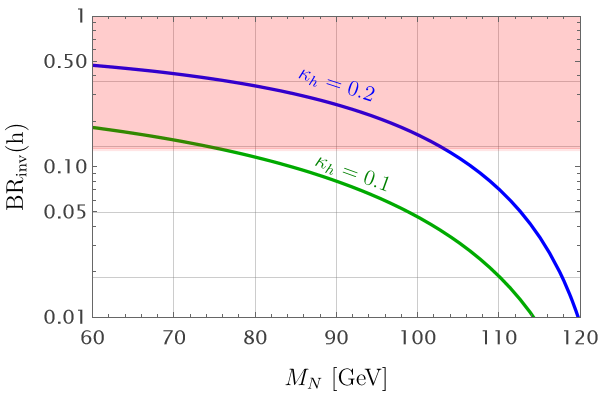}
        \caption{}
        \label{subfig:hinv}
        \end{subfigure}
    \begin{subfigure}[b]{0.48\textwidth}
        \centering
        \includegraphics[width=\linewidth]{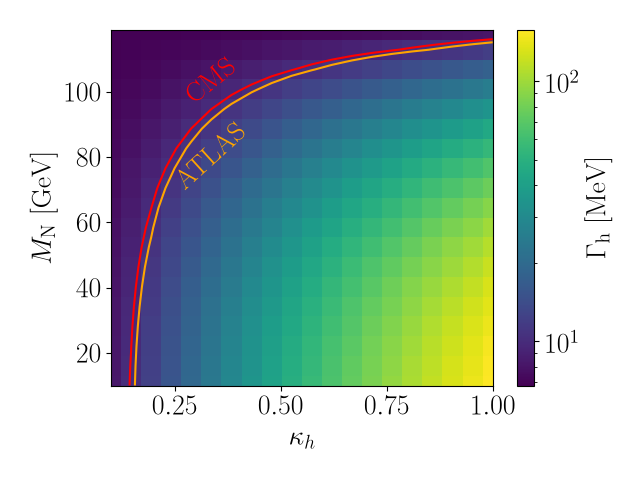}
        \caption{}
        \label{subfig:hwidth}
    \end{subfigure}
    \caption{(a) Branching ratio of the SM-like Higgs decay to the the right-handed neutrino $N$, as a function of \MN.
      For scenarios in which the heavy neutrino is collider-stable, the red shaded region is excluded by the
      CMS measurements of
      invisible Higgs decays~\cite{CMS:2026nce}. (b) The effect of \MN and $\kappa_h$ on the total width of the Higgs.
      The orange, ATLAS \cite{ATLAS_higgs_width}, and red, CMS \cite{CMS_higgs_width}, lines are the 95\% CL
      upper-bound on $\ensuremath{\Gamma_{\mathrm{h}}}$. For both we assume that the three generations of right-handed neutrinos are
      degenerate in mass.}
 \end{figure}

 In this unification framework, if kinematically allowed, the SM like Higgs, $h$, can decay to  $N_i \nu_i$.
 In our convention, the coupling of the SM like higgs and heavy and active neutrinos can be written as 
 \begin{eqnarray}
     C_{N_i\nu_i}^h = \frac{\kappa_h}{4 v} m_{u_i}.
 \end{eqnarray}

 \begin{figure}[t]
    \centering
    \begin{subfigure}[b]{0.49\textwidth}
        \centering
        \includegraphics[width=\textwidth]{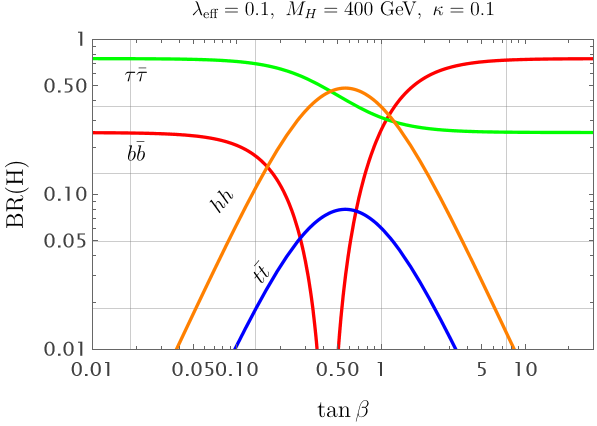}
        \caption{}
        \label{subfig:H1}
    \end{subfigure}  
    \begin{subfigure}[b]{0.49\textwidth}
        \centering
        \includegraphics[width=\textwidth]{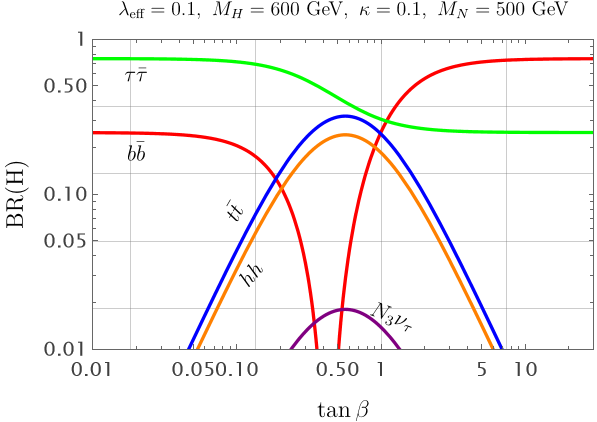}
        \caption{}
        \label{subfig:H2}
    \end{subfigure}          
    \begin{subfigure}[b]{0.49\textwidth}
        \centering
        \includegraphics[width=\textwidth]{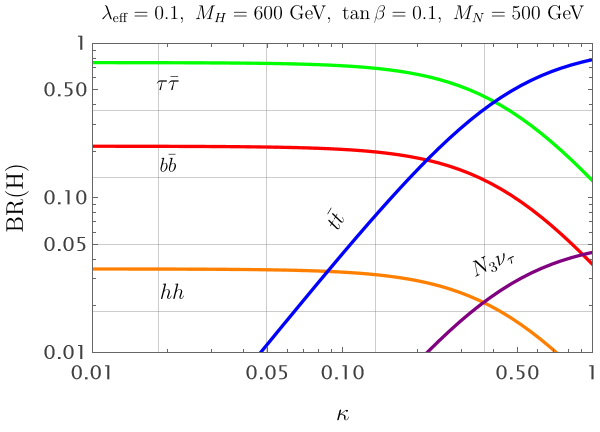}
        \caption{}
        \label{subfig:H3}
    \end{subfigure} 
    \begin{subfigure}[b]{0.49\textwidth}
        \centering
        \includegraphics[width=\textwidth]{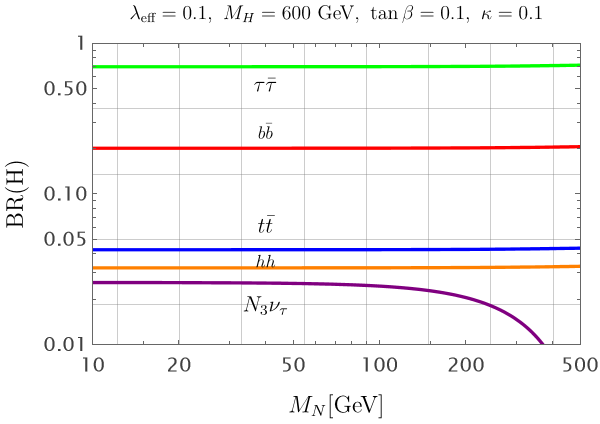}
        \caption{}
        \label{subfig:H4}
    \end{subfigure} 
    \begin{subfigure}[b]{0.55\textwidth}
        \centering
        \includegraphics[width=\textwidth]{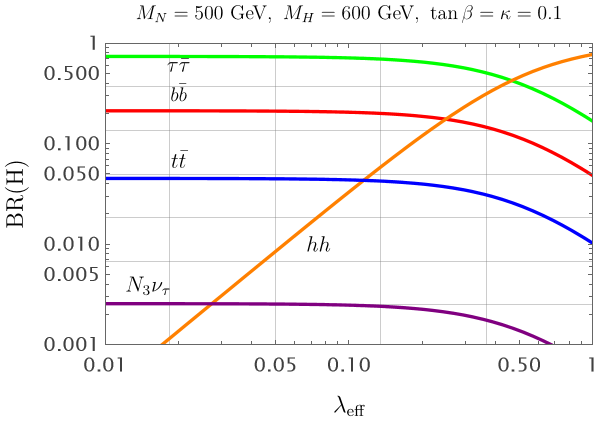}
        \caption{}
        \label{subfig:H5}
    \end{subfigure} 
    \caption{Branching ratios of the CP-even Higgs, H decay.
      a) Scenario where the decay into right-handed neutrinos is kinematically forbidden.
      b) Scenario where the decay into right-handed neutrinos is kinematically allowed.
      In both cases, we assume $\leff = \kappa = 0.1, \MN=500$ GeV.
      c) Scenario with $\tanb = \leff = 0.1 $, $\MN =500 $ GeV
      and varying $\kappa$.
      d) Scenario with $\tanb = \leff = \kappa = 0.1 $  and varying $\MN$.
      e) Scenario with $\tanb = \kappa = 0.1, \MN = 500$ GeV  and varying \leff. }
    \label{fig:CPevenHdecay}
 \end{figure}
 When the new Higgs bosons and leptoquarks are heavy, and when the mixing angles between the SM neutrinos and heavy neutrinos
 are very small, the right-handed neutrinos can become long-lived,
 making this decay channel invisible.
 In Fig.~\ref{subfig:hinv}, we show the branching ratio of this invisible decay as a function of right-handed neutrino
 mass for different values of $\kappa_h$ assuming the right handed neutrinos are long lived.
 One can see that for values of \MN below 100~GeV or so, the value of $\kappa_h$ has to be small to satisfy the
 experimental bounds.
 Note that if the new Higgs bosons are light, the decay length of the right handed
 neutrinos may become shorter than the detector size, making this channel detectable through the decays of
 the right-handed neutrinos.
 On the other hand, when the right handed neutrinos are heavier than the SM like Higgs h, this decay channel is
 kinematically forbidden.
 In these scenarios, the constraints on $\kappa_h$ from this process no longer apply.

 The ATLAS and CMS measurements of the SM Higgs width are also relevant here, since at large $\kappa_h$ and small \MN
 the width of the SM Higgs could grow significantly. This is shown in Fig.~\ref{subfig:hwidth}, where in the bottom
 right region of the plot, below the red and orange lines, the SM Higgs width given by this model would be
 in conflict with data. Notice that the top left section remains allowed.
 \section{NEW HIGGS BOSONS DECAYS}
   \label{sec:bsmhiggs}
%
\begin{figure}[hbt]
    \centering
    \begin{subfigure}[b]{0.49\textwidth}
        \centering
        \includegraphics[width=\textwidth]{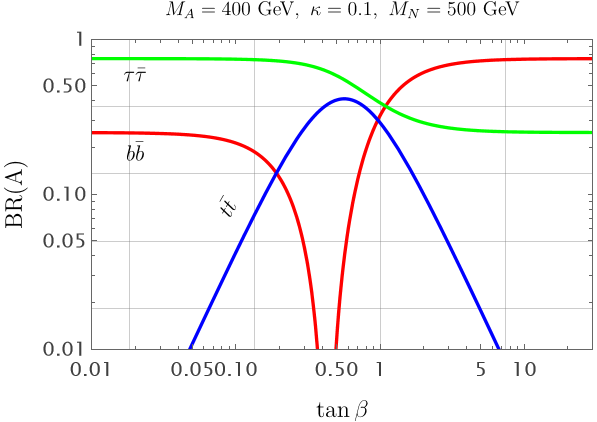}
        \caption{}
        \label{subfig:A1}
    \end{subfigure}  
    \begin{subfigure}[b]{0.49\textwidth}
        \centering
        \includegraphics[width=\textwidth]{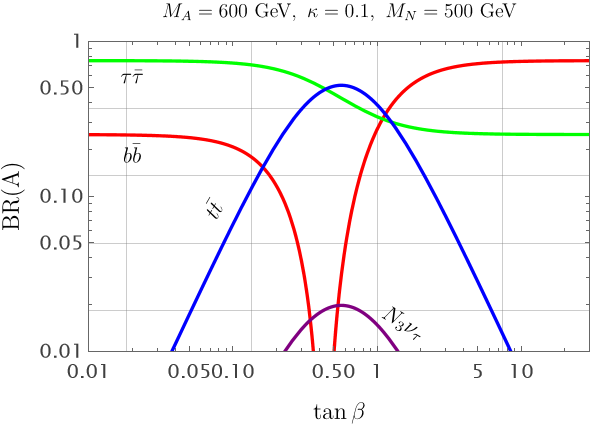}
        \caption{}
        \label{subfig:A2}
    \end{subfigure}          
    \begin{subfigure}[b]{0.49\textwidth}
        \centering
        \includegraphics[width=\textwidth]{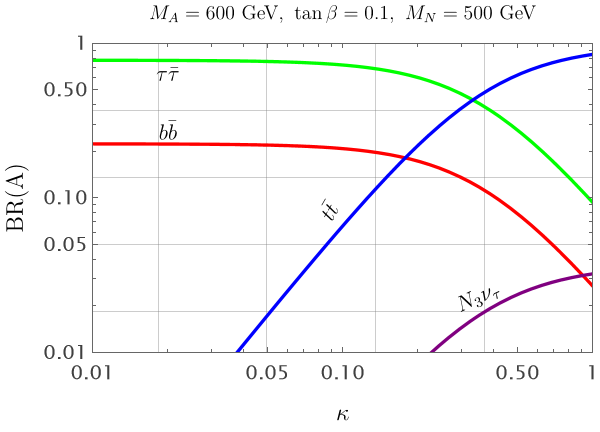}
        \caption{}
        \label{subfig:A3}
    \end{subfigure} 
    \begin{subfigure}[b]{0.49\textwidth}
        \centering
        \includegraphics[width=\textwidth]{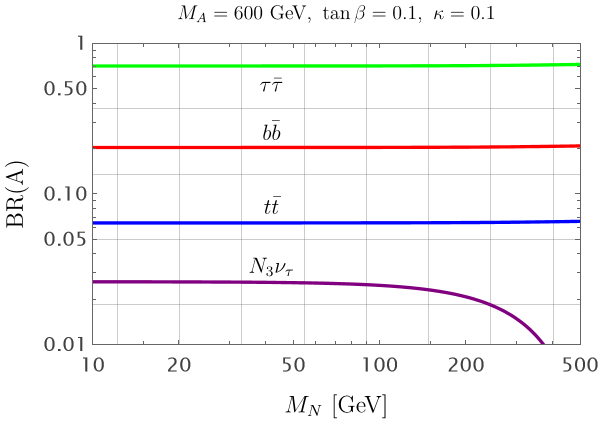}
        \caption{}
        \label{subfig:A4}
    \end{subfigure} 
    \caption{Branching ratios of the CP-odd Higgs, A, decay.
      a) Scenario where the decay into right-handed neutrinos is kinematically forbidden.
      b)  Scenario where the decay into right-handed neutrinos is kinematically allowed.
      In both cases, we assume $\kappa=0.1$, $\MN=500$ GeV.
      c) Scenario with $\tanb = 0.1 $, $\MN =500 $ GeV and varying $\kappa$.
      d) Scenario with $\tanb = \kappa=0.1$  and varying $\MN$.}
    \label{fig:CPoddAdecay}
 \end{figure}
This quark-lepton unification framework predicts unique relations between different decay modes of the CP-even and CP-odd Higgs bosons. The dominant decay channels of the new Higgs bosons are third-generation SM fermions, while decays into the heavy right-handed neutrino occur when kinematically allowed. The heavy right-handed neutrino can then decay into SM fermions through interactions involving an off-shell scalar leptoquark or the new Higgs bosons. In the alignment limit, $\sin(\beta-\alpha)\to1$, the parameter entering in the Yukawa coupling between the H and the charged leptons and down quarks can be written as 
\begin{eqnarray}
    C_{dd}^H & = & (3 \tanb - \cot\beta) \frac{m_d^{diag}}{4 v} + (\tanb + \cot \beta ) \frac{m_e^{diag}}{4 v}, 
    \label{eq:hddint}
    \\
    C^H_{ee} &= & (\tanb -3\cot{\beta})\frac{m_e^{diag}}{4v}+ 3(\tanb+\cot{\beta})\frac{  m_d^{diag}}{4v}.
\end{eqnarray}
Given the flexibility in choosing the Yukawa couplings $Y_1$, $Y_2$ and the mixing matrices, we parametrize the coupling of up quarks with a free parameter $\kappa$ as follows 
\begin{eqnarray}
    C_{uu}^{H} =C^{H}_{N\nu} = \frac{\kappa}{4 v} m_u^{\text{diag}}. \hspace{0.2 cm}
    \label{eq:Huufeyn}
\end{eqnarray}
As can be seen from Eqs.(\ref{eq:hddint}-\ref{eq:Huufeyn}), the couplings are proportional to the fermion masses, implying that decays into the 3rd generation quarks and leptons are dominant. For details, see appendix \ref{appendix}. If kinematically allowed, the new neutral CP-even Higgs, H, mainly decays to 
\begin{eqnarray}
    H \to \bar{t}t, \ \bar{b}b, \ \bar{\tau}\tau, \  hh, \ N \nu.
\end{eqnarray}
\begin{figure}[t]
    \centering
    \begin{subfigure}[b]{0.49\textwidth}
        \centering
        \includegraphics[width=\textwidth]{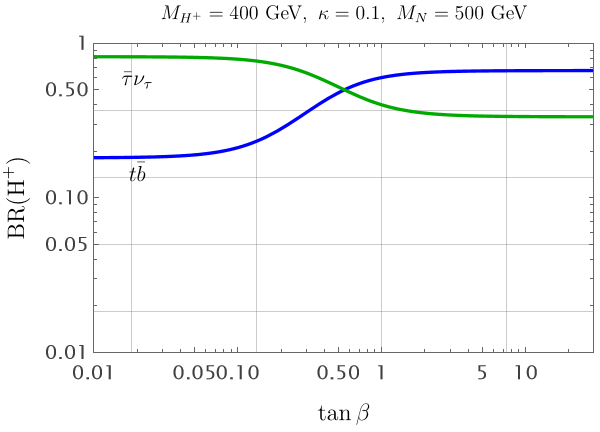}
        \caption{}
        \label{subfig:C1}
    \end{subfigure}  
    \begin{subfigure}[b]{0.49\textwidth}
        \centering
        \includegraphics[width=\textwidth]{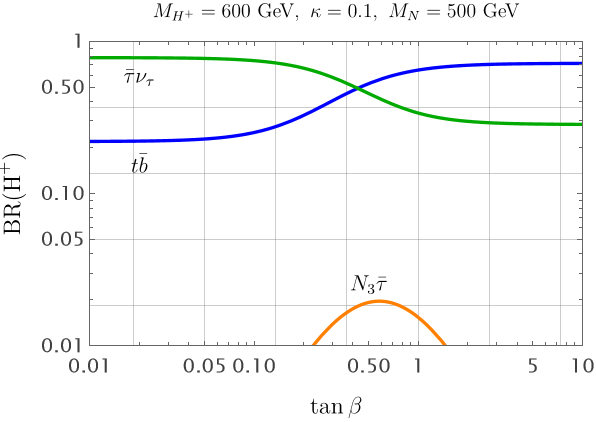}
        \caption{}
        \label{subfig:C2}
    \end{subfigure}          
    \begin{subfigure}[b]{0.49\textwidth}
        \centering
        \includegraphics[width=\textwidth]{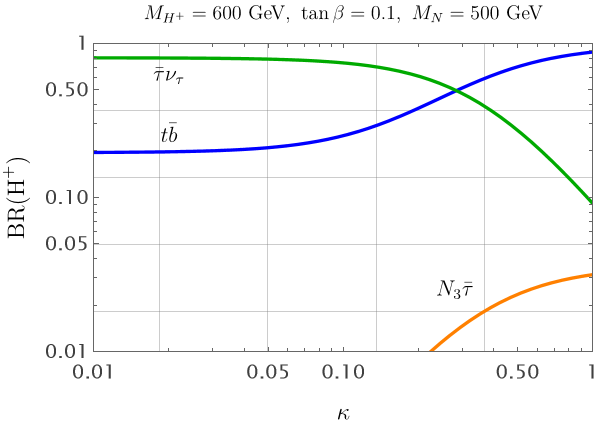}
        \caption{}
        \label{subfig:C3}
    \end{subfigure} 
    \begin{subfigure}[b]{0.49\textwidth}
        \centering
        \includegraphics[width=\textwidth]{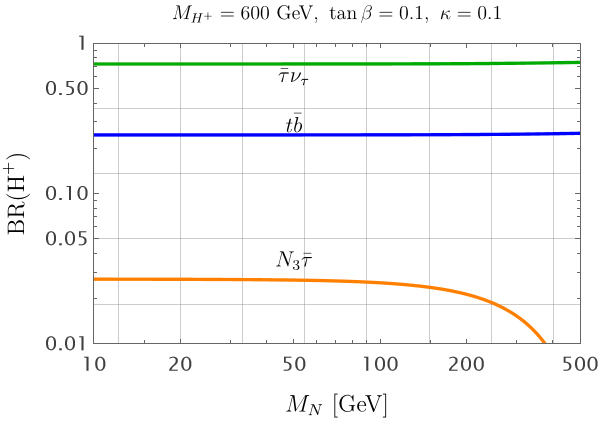}
        \caption{}
        \label{subfig:C4}
    \end{subfigure} 
    \caption{Branching ratios of the charged Higgs, $H^+$ decay.
      a) Scenario where the decay into right-handed neutrinos is kinematically forbidden.
      b)  Scenario where the decay into right-handed neutrinos is kinematically allowed.
      In both cases, we assume $\kappa=0.1$ , \MN=500~GeV.
      c) Scenario with $\tanb = 0.1 $, $\MN =500 $ GeV and varying $\kappa$.
      d) Scenario with $\tanb = \kappa = 0.1$  and varying \MN. }
    \label{fig:ChargedHdecay}
 \end{figure}
 When there is no flavor violation,  in the limits $\tanb \ll 1 $ and $\tanb \gg 1$, the quark–lepton unification framework predicts unique relations for the decay of the heavy CP-even Higgs boson. In particular, for  $\tanb \ll 1$, it leads to,
 \begin{eqnarray}
    \Gamma(H \to \bar{\tau}\tau) = 3 \Gamma(H \to  \bar{b} b),
\end{eqnarray}
while in the opposite regime, when $\tanb \gg 1 $, one obtains the following relation.
\begin{eqnarray}
    \Gamma(H \to \bar{\tau}\tau) = \frac{1}{3} \Gamma(H \to  \bar{b}b).
\end{eqnarray}
In Fig.~\ref{fig:CPevenHdecay} we show the branching ratios of the CP-even Higgs decays for different scenarios.
Fig.~\ref{subfig:H1} shows the branching ratios as a function of $\tan\beta$ assuming
$\leff=\kappa=0.1, \MH =400$ GeV.
As expected, in the small $\tan \beta$ region, the $\tau \tau$ decay dominates, while in the large $\tan\beta$ region,
the $b \bar{b}$  channel becomes the dominant decay mode. However, in the intermediate region, where $ \tan\beta$ is $\mathcal{O}(1)$, the decay into the two SM like higgs suppresses the fermionic decay channels. Fig.~\ref{subfig:H2} shows the branching ratios when the decay into right-handed neutrinos is kinematically allowed. 

The branching fractions of the CP-even Higgs decays also depend on the value of $\kappa$, and this dependency is illustrated in Fig.~\ref{subfig:H3} assuming $\tanb=0.1$. As the value of $\kappa$ increases, the decay branching ratios change significantly, with the $t\bar{t}$ channel becoming increasingly important. We also show in Fig.~\ref{subfig:H4} the branching ratio dependence on right-handed neutrino masses. For relatively light right-handed neutrinos, the decay channel $H \to N_i\nu_i$ can have a large branching fraction. To show the importance of the Higgs decays into two SM-like Higgses, in Fig.~\ref{subfig:H5} we have the numerical predictions for the branching ratios vs. $\leff$. In this case, when $\leff$ is large the channel with two SM-like Higgs dominates.  

The interaction terms for the CP-odd Higgs, $A$, can be expressed in a similar way 
 \begin{eqnarray}
     C_{dd}^A & = &(3 \tanb - \cot\beta) \frac{m_d^{diag}}{4 v} + (\tanb + \cot \beta ) \frac{ m_e^{diag}}{4 v} \label{eq:Addint},\\
      C^A_{ee} &=& (\tanb -3\cot{\beta})\frac{m_e^{diag}}{4v}+ 3(\tanb+\cot{\beta})\frac{ m_d^{diag}}{4v},\\
      C_{uu}^{A} &=& C^{A}_{N\nu} = \frac{\kappa}{4 v} m_u^{\text{diag}}.
      \label{eq:Aeeint}
 \end{eqnarray}
 Therefore, as discussed above, the decays into the third generation dominate. The CP-odd Higgs A mostly decays to 
\begin{eqnarray}
    A \to \bar{t}t , \ \bar{b} b, \ \bar{\tau} \tau,N \nu. 
\end{eqnarray}
 If the right-handed neutrinos N are heavier than the CP-odd Higgs A, the decay $A \to N \nu$ is kinematically forbidden. This theory predicts the following  unique relations for the decays of CP-odd Higgs when we neglect the quark masses:
\begin{eqnarray}
    \Gamma(A \to \bar{\tau}\tau) = 3 \Gamma(A \to  \bar{b}b) , \hspace{1 cm} \tanb \ll 1, \\
    \Gamma(A \to \bar{\tau}\tau) = \frac{1}{3} \Gamma(A \to  \bar{b}b) , \hspace{1 cm} \tanb \gg 1.
\end{eqnarray}
In Fig.~\ref{fig:CPoddAdecay}, we illustrate the branching ratios of the different decay modes of the CP-odd Higgs boson. It is important to mention that the overall decay behavior into SM fermions is qualitatively similar to that of the CP-even Higgs boson, H. However, a key distinction is that the CP-odd Higgs boson does not decay into a pair of SM like Higgs bosons.
Notice that in this case one also has the unique relation between the decays into bottom quarks and tau leptons.

Similarly, the interaction between the charged Higgs boson and the leptons can be written as 
\begin{eqnarray}
     C_{\nu e}^{H^+}&=&  3(\tanb+\cot{\beta})\frac{  m_d^{diag}}{2 \sqrt{2}v}+ (\tanb-3\cot{\beta})\frac{ m_e^{diag}}{2\sqrt{2}v}.
\end{eqnarray}
As the interaction is proportional to the masses of the fermions, the dominant decay channels are in third-generation fermions.
The  new charged Higgs boson  mostly decays  to 
\begin{eqnarray}
    H^+ \to \bar{\tau} \nu_\tau , \ t\bar{b} , \bar{\tau}N.
\end{eqnarray}
In Fig.~\ref{fig:ChargedHdecay} we illustrate the branching ratios of the new charged Higgs, $H^+$, decay assuming different benchmark scenarios. When $\kappa$ is small, in the low $ \tan\beta$ region, the decay into $\tau \nu_\tau$ dominates, while in the large $ \tan\beta$ region, the $t\bar{b}$ decay mode dominates. When kinematically allowed, the charged Higgs can also decay to a right-handed neutrino and a SM charged lepton, and the branching ratio can be large for light right-handed neutrinos.  
\section{LHC CONSTRAINTS}
\label{sec:contur}
\begin{figure}[t]
   \centering
    \begin{subfigure}[h]{0.48\textwidth}
        \centering
        \includegraphics[width=\textwidth]{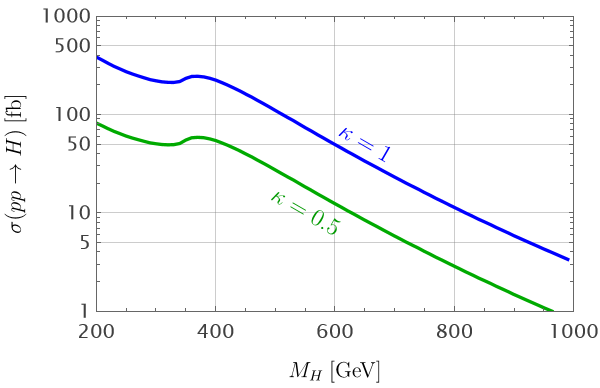}
        \caption{}
        \label{subfig:crossH}
    \end{subfigure}
    \begin{subfigure}[h]{0.48\textwidth}
        \centering
        \includegraphics[width=\textwidth]{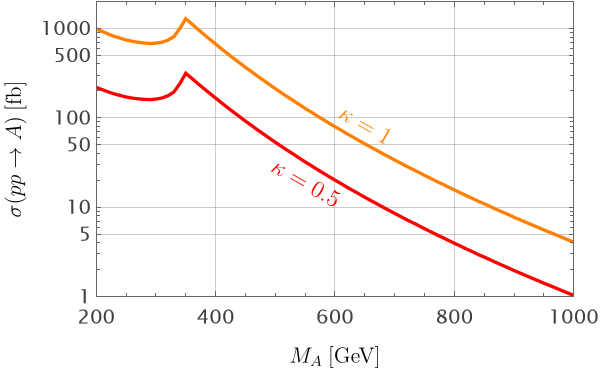}
       \caption{}
       \label{subfig:crossA}
    \end{subfigure}
    \begin{subfigure}[h]{0.48\textwidth}
        \centering
        \includegraphics[width=\textwidth]{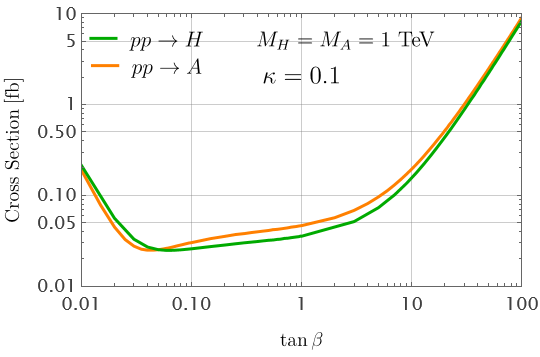}
       \caption{}
       \label{subfig:crosstB}
    \end{subfigure}
    \caption{Production cross section for a) CP even Higgs, H, and b) CP-odd Higgs, A at $\sqrt{s}=13$ TeV, as a
      function of Higgs mass for different values of $\kappa$ when $\tanb =1$. c) Production cross section for both
      CP-even and CP-odd Higgs as a function of $ \ \tan\beta$ when $M_H=M_A=1$ TeV and $\kappa=0.1$.
      The cross section was computed with our UFO model and  \madgraph ~\cite{Alwall:2014hca}. }
\label{fig:crossHiggs}
\end{figure}

The final states generated by the production and decay processes discussed in the previous section could in principle
leave their imprint on several of the fiducial cross sections which have been measured already at the LHC, or which
are likely to be measured in future. In this section we use the \contur~\cite{Butterworth:2016sqg,Buckley:2021neu}
application to study the extent to which the model is already constrained by these measurements - all of which are,
to date, consistent with the SM.

To calculate the production cross sections and decay modes of the model, we implemented the theory in
FeynRules~\cite{Alloul:2013bka} and exported it in  UFO format. Events are simulated using
\herwig \cite{Bellm:2015jjp}, based upon the UFO~\cite{Degrande:2011ua,Degrande:2014vpa} files of the theory.
The simulated events are then passed through \rivet~4~\cite{Buckley:2010ar,Bierlich:2024vqo} and analysed with
\contur~3.1~\cite{CONTUR:2025yis}.
This involves injecting the BSM Higgs contribution on top of the SM predictions for a wide range
of LHC measurements, and evaluating the ratio between the likelihood evaluated for the SM alone given the data,
and the likelihood for SM+BSM contributions.
In doing this, the data are divided into ``analysis pools'', based upon experiment, run period and final state, so that each pool
can be treated as statistically independent of the others\footnote{This was shown in
Ref.~\cite{LHCREIWGOpenEventGenerationTaskForce:2026dvu} to capture the dominant correlation pattern.}.
Only the most sensitive measurement in a given pool
is combined into the overall likelihood.
The \cls method~\cite{Read:2002hq} is used to determine an exclusion limit;
the expected exclusion limit is also determined, by setting the data to the SM values but preserving its uncertainties.
Also shown is a crude estimate of the potential reach of the high-luminosity LHC, obtained by scaling the experimental uncertainties
of the current measurements according to a target luminosity of 3000~fb$^{-1}$. This is a conservative estimated since it does not
take into account any extension of the measurements into new phase space.

In what follows, we will discuss each of the new Higgs bosons in turn, investigating how the parameters $\tanb, \leff$
and $\kappa$ affect their observability at the LHC, as well as, where relevant, the influence of the heavy neutrino.
In all cases will assume the leptoquarks are at sufficiently high mass that their signatures are not
visible\footnote{See Ref.\cite{Butterworth:2025ttw} for a study of the phenomenology of lower mass leptoquarks in this
model.}.
The dominant production mechanism for both the CP-even and CP-odd Higgs bosons is gluon fusion, while the new
charged Higgs is mainly produced via electroweak interactions. Then, the main pair production mechanisms are: 
\begin{eqnarray}
    g \  g \to H /A  , \hspace{1 cm} p \ p  \to \gamma ,Z \to H^\pm H^\mp.
\end{eqnarray}
Because the new charged Higgs boson is mainly produced via electroweak interactions,
its pair-production cross section is relatively small, although single production via $g b \rightarrow H^\pm t$
can be quite large at high \tanb.

In Fig.~\ref{fig:crossHiggs}, we show the production cross section at the LHC, for proton–proton collisions at a
center-of-mass energy of 13 TeV, for both CP-even and CP-odd Higgs as a function of the corresponding Higgs mass
and for different values of $\kappa$.
In Fig.~\ref{subfig:crossH} we illustrate the gluon fusion cross section for the CP-even Higgs for
$\kappa =1 \ \text{and }  0.5$ and $\tanb=1$.
The production cross section for the CP-odd Higgs boson, considering the same benchmark scenarios,
is shown in Fig.~\ref{subfig:crossA}.
Note that in both cases, the production cross section is of the order of several femtobarns, even hundreds of
femtobarns, and increases with larger values of $\kappa$ and $\tan\beta$.
\begin{figure}[htbp]
  \centering
  \begin{subfigure}[b]{\textwidth}
    \includegraphics[width=0.9\textwidth]{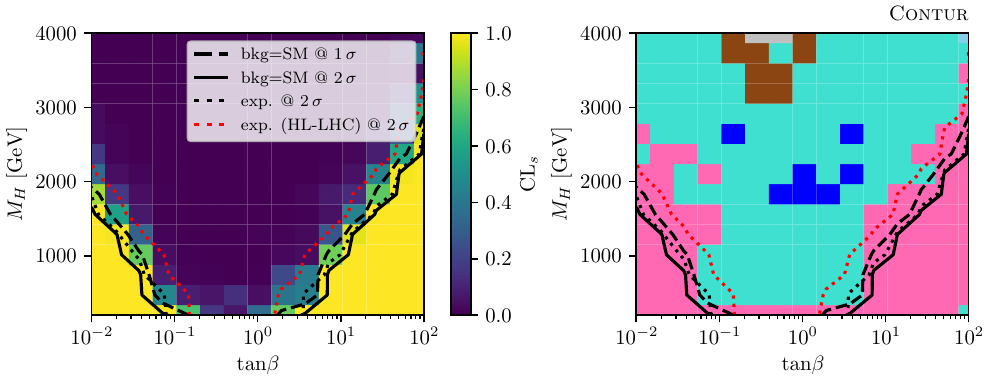}
    \caption{Scan over CP-even Higgs mass and \tanb with $\kappa=\leff=0.1$.}
    \label{fig:mhbeta}
  \end{subfigure}  
  \centering
  \begin{subfigure}[b]{\textwidth}
    \centering
    \includegraphics[width=0.9\textwidth]{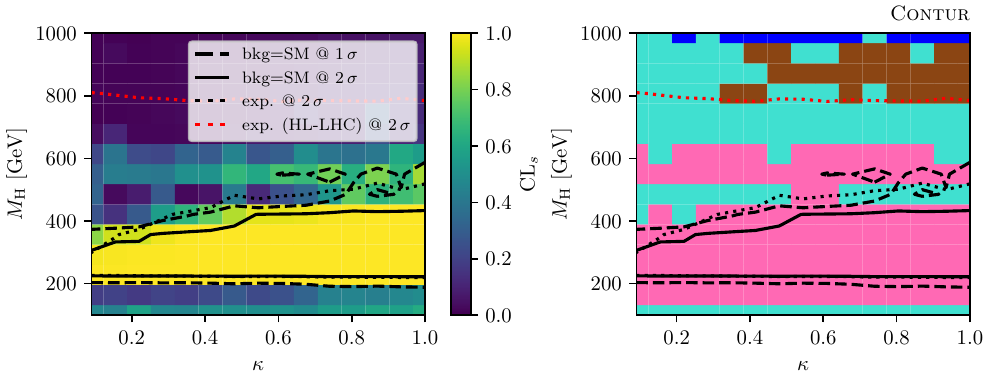}
    \caption{Scan over CP-even Higgs mass and $\kappa$ with $\leff=\tanb=0.1$.}
    \label{fig:mhk}
  \end{subfigure}
  \centering
  \begin{subfigure}[b]{\textwidth}
    \includegraphics[width=0.9\linewidth]{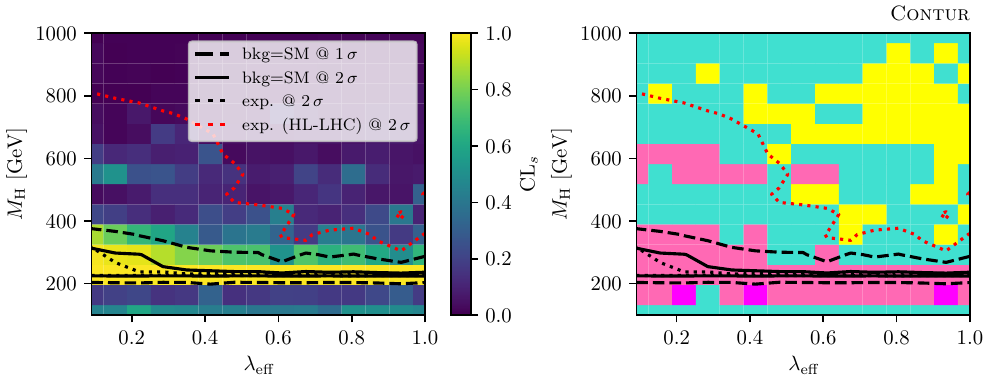}
    \caption{Scan over CP-even Higgs mass and \leff with $\kappa=\tanb=0.1$.}
    \label{fig:mhleff}
  \end{subfigure}
  \begin{tabular}{llll}
    \swatch{magenta}~4$\ell$ \cite{ATLAS:2021kog} & 
    \swatch{hotpink}~$\tau^+\tau^-$ \cite{ATLAS:2025oiy} & 
    \swatch{turquoise}~$\ell_1\ell_2$+\MET{}+jet \cite{ATLAS:2023gsl,ATLAS:2024aht}  &
    \swatch{blue}~$\ell$+\MET{}+jet \\
    \swatch{yellow}~$h\to\gamma\gamma$ (right plot only) &
    \swatch{saddlebrown}~hadronic $t\bar{t}$  & 
    \swatch{silver}~jets
  \end{tabular}
  \caption{Scans over the CP-even higgs mass $\MH$ and other parameters of the model. In all cases,
    $\MN$=500 GeV, $V_{mix}=10^{-4}$ and the leptoquarks, CP-odd and charged Higgs bosons are decoupled.
    The solid (dashed) black line indicates the 95\% (68\%) exclusion. The dotted black (red) line shows
    the expected 95\% exclusion for the LHC (HL-LHC). The most sensitive  analysis pool at each point in a parameter scan
    is indicated in the plots by colour coding, with the relevant references given in the plot legend.
  }
  \label{fig:MHscans1}
\end{figure}
In Fig.~\ref{subfig:crosstB} the cross section is shown as a function of \tanb for $\MA = \MH = 1$~TeV and $\kappa = 1$.
It can be seen that the cross section rises at high \tanb and low \tanb (i.e. high $\cot\beta$) as expected from
the behaviour of the couplings to quarks (appearing in the loop coupling to gluons) shown in
Eq.~\ref{eq:hddint} and Eq.~\ref{eq:Addint}. In between the extremes of large and small \tanb, the cross section goes
through a minimum, where one has a cancellation between the top and bottom quark contributions.
\subsection{CP-even Higgs}
\label{sec:cpeven}
\begin{figure}[t] \includegraphics[width=0.9\linewidth]{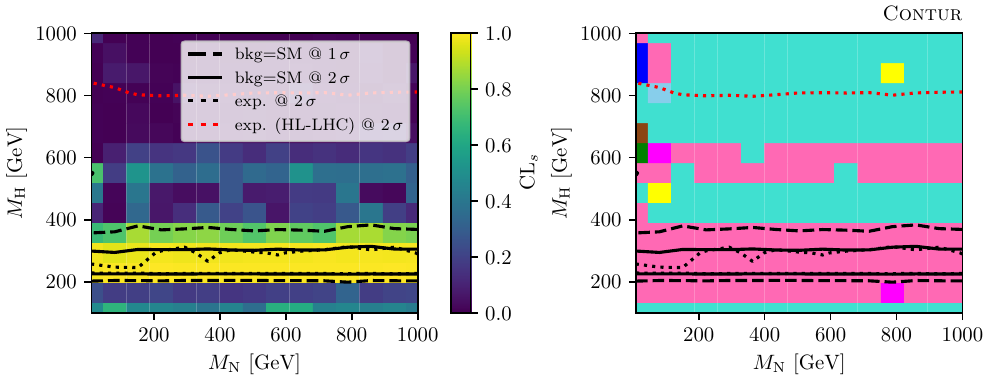}
  \label{fig:mhmn}
  \caption{Scan over $\MH$ and $\MN$ for $\kappa=\leff=\tanb=0.1$ and $V_{mix}=10^{-4}$.
    The CP-odd and charged Higgs bosons are decoupled and leptoquarks only play a potential
    role through the decays of $N$.
    Other details as Fig.~\ref{fig:MHscans1}.
  }
  \begin{tabular}{llll}
    \swatch{turquoise}~$\ell_1\ell_2$+\MET{}+jet \cite{ATLAS:2023gsl,ATLAS:2024aht} & 
    \swatch{magenta}~4$\ell$ \cite{ATLAS:2021kog} & 
    \swatch{blue}~$\ell$+\MET{}+jet  & 
    \swatch{hotpink}~$\tau^+\tau^-$ \cite{ATLAS:2025oiy} \\ 
    \swatch{green}~\MET{}+jet  & 
    \swatch{saddlebrown}~hadronic $t\bar{t}$  & 
    \swatch{skyblue}~$\ell_1\ell_2$+\MET{}  & 
    \swatch{yellow}~$h\to\gamma\gamma$ (right plot only) \\ 
    \end{tabular}
  \label{fig:MHscans2}
\end{figure}

The sensitivity of the LHC measurements in \contur to the CP-even Higgs is shown in Fig.~\ref{fig:MHscans1} and
Fig.~\ref{fig:MHscans2}.
In Fig.\ref{fig:mhbeta} it can be seen that at both small and large \tanb, the model is excluded up to fairly high
values of \MH -- nearly 2~TeV at $\tanb \approx 10^{-2}$ and nearly 3~TeV at $\tanb \approx 10^{2}$
when $\kappa = \leff = 0.1$.
This exclusion comes mainly from the inclusive measurement of di-$\tau$ production~\cite{ATLAS:2025oiy}, since as
shown in Figs.~\ref{subfig:H1} and \ref{subfig:H2}, this channel dominates in these regions (apart from decays to $b$-quarks, which are harder to identify above the large QCD background).

In the middle region where \tanb is closer to one, decays to SM Higgs bosons and top quarks dominate; again
these result mainly in $b$-jets and the sensitivity is thus reduced. The model is allowed by the measurements
in \contur. However, di-higgs searches provide upper bounds on the cross-section for resonant pair production of SM
higgs bosons. Current limits at 95\% c.l. allow a maximum cross section for SM Higgs pair production
ranging from around 10-300 fb as resonant
mass (in our case \MH or \MA) ranges from around 600 to 250 GeV~\cite{ATLAS:2023vdy,CMS:2026mwf}. Referring to
Fig.~\ref{fig:crossHiggs}, it can be see that at $\tanb \approx 1$ these limits can exclude high values
of $\kappa$ for much of the \MH range considered, for \leff = 0.1 or higher.
Nevertheless, for low enough $\kappa$, much of the \MA, \MH and \tanb parameter space will remain allowed.
The combination of ATLAS and CMS results \cite{CMS:2026nuu} for non-resonant di-Higgs production
gives an observed limit at 95\% CL corresponding to 73 fb, 2.5x the SM $\sigma_{ggF}+\sigma_{VBF}$ prediction,
leading to similar conclusions.
Note that as shown in Fig.~\ref{subfig:H5}, lowering \leff below our benchmark of 0.1 will further suppress
decays to SM Higgs pairs, thus evading these limits, but correspondingly
strengthening the sensitivity of the di-$\tau$ and \ttbar measurements.


As expected from Fig.~\ref{subfig:H3}, the dependence of the sensitivity on $\kappa$ when $\tanb = \leff = 0.1$ is mild,
as shown in Fig.~\ref{fig:mhk}. \MH values below 200~\GeV \ are allowed, and the there an exclusion band above this.
The upper limit of this band grows from about 300~\GeV \ to above 400~GeV as $\kappa$ grows.
Again, the di-$\tau$ measurement is the most sensitive, with additional contributions from fully-leptonics \ttbar measurements.

The story is similar when \leff is varied, for $\kappa = \tanb = 0.1$
(Fig.~\ref{fig:mhleff}) except that the excluded mass region is now just 200~\GeV $< \MH < 400$~\GeV.
Note that as shown in Fig.\ref{subfig:H5}, at high values of \leff, the branching ratio to SM Higgs pairs becomes
large, and the diHiggs limits discussed above, which are not included in \contur, will have an impact.
The phenomenology can in principle depend upon \MN, if it is low enough that the $H$ decays to $N$ shown in
Fig.\ref{subfig:H3} are kinematically allowed. Notice that $N$ decays via the leptoquark and Higgses into SM particles. Nevertheless, we see from Fig.~\ref{fig:MHscans2} that this dependence is weak.

\begin{figure}[t]
    \centering
    \begin{subfigure}[h]{0.48\textwidth}
        \centering
        \includegraphics[width=\textwidth]{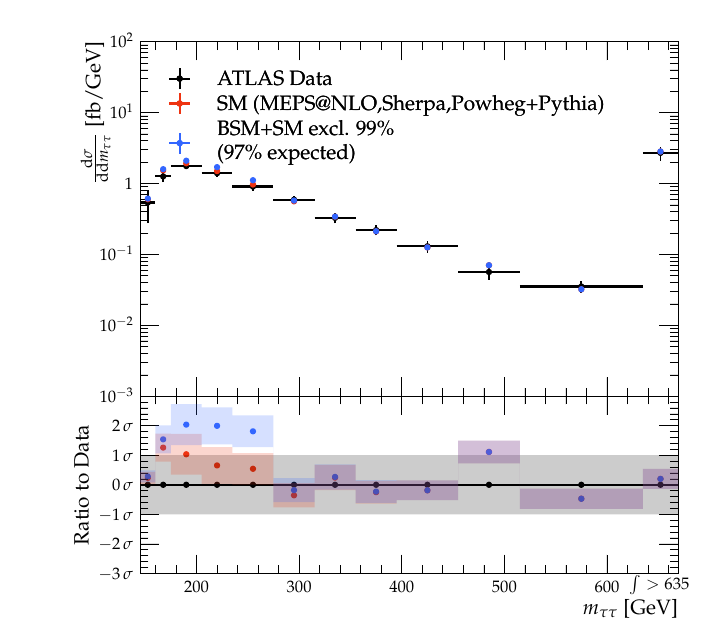}
        \caption{}
        \label{rivet:high-mass-low-lambda}
    \end{subfigure}
    \begin{subfigure}[h]{0.48\textwidth}
    \includegraphics[width=\textwidth]{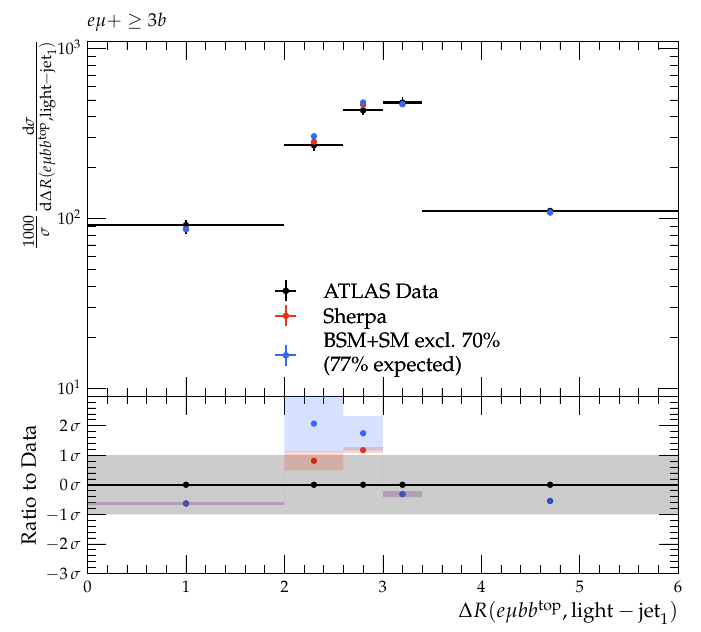}
    \caption{}
    \label{rivet:rivet:high-mass-high-lambda}
    \end{subfigure}
    \caption{Difference between SM and SM+BSM for the most excluding distributions for two points from \ref{fig:mhleff} (a) $\MH=289$ GeV and $\leff=0.01$, most excluded by $\tau^{+}\tau^{-}$ and (b) $\MH=289$ GeV and $\leff=1$, most excluded by $\ell_1\ell_2+\MET+jet$.
    \label{fig:leffhistos}}
\end{figure}

It is notable that significantly increased sensitivity to the model is expected from the HL-LHC,
show by the red dotted line
in the Figures. In particular, for $\tanb= 0.1$, the mass range extends up to around 800~\GeV.
This is driven not only by the
$\tau$ measurement, but also by increasingly precise \ttbar measurements. As an example, the most sensitive measurements
for two points in Fig.~\ref{fig:mhleff} are shown in Fig.~\ref{fig:leffhistos}. Both points are at relatively low
\MH. Fig.\ref{rivet:high-mass-low-lambda}, at \leff=0.01, is excluded already by the di-$\tau$ measurement, and
the excess in the prediction over
data can be clearly seen for visible di-$\tau$ masses up to that value. Note that the measurement is of
the visible mass, i.e.
excluding the $\nu_\tau$, and so the excess is expected to show below \MH.
Figure~\ref{rivet:rivet:high-mass-high-lambda} shows
a point at \leff = 1, where the model is not currently excluded but where the HL-LHC is expected to be sensitive.
For that point, the
current most sensitive measurement is a recent fully-leptonic \ttbar measurement from ATLAS~\cite{ATLAS:2024aht},
in particular the
radial separation of the $e\mu b \bar{b}$ top system and hardest additional light (or charm) jet is the event.
%

\begin{figure}[hbt]
  \begin{subfigure}[b]{\textwidth}
    \centering
    \includegraphics[width=0.9\textwidth]{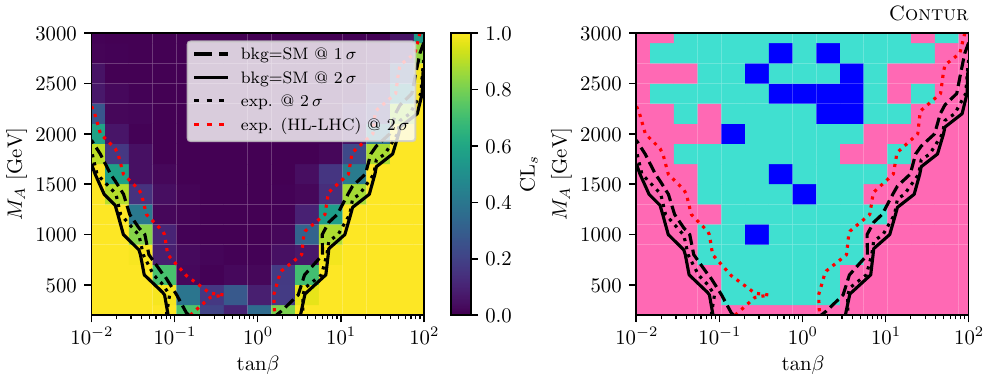}
    \caption{Scan over CP-odd Higgs mass and \tanb with $\kappa=0.1$.
    }
    \label{fig:mabeta}
  \end{subfigure}  
  \centering
  \begin{subfigure}[b]{\textwidth}
    \centering
    \includegraphics[width=0.9\textwidth]{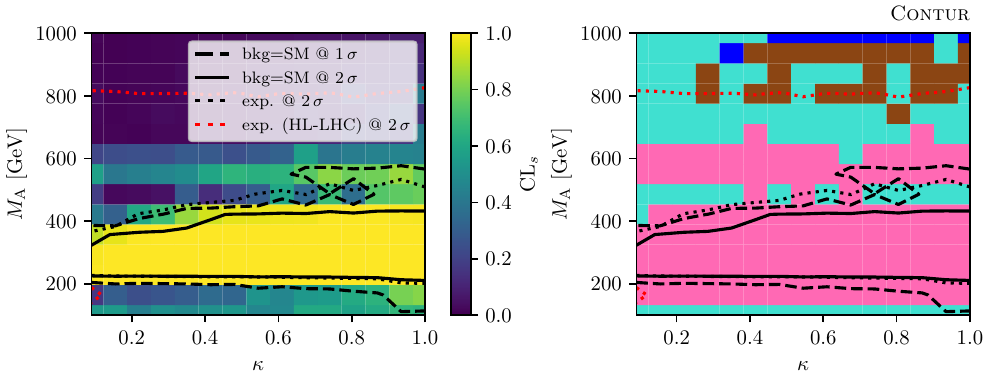}
    \caption{Scan over CP-odd Higgs mass and $\kappa$ with $\tanb=0.1$.}
    \label{fig:mak}
  \end{subfigure}
  \centering
  \begin{tabular}{llll}
    \swatch{saddlebrown}~hadronic $t\bar{t}$ & 
    \swatch{hotpink}~$\tau^+\tau^-$ \cite{ATLAS:2025oiy} & 
    \swatch{blue}~$\ell$+\MET{}+jet & 
    \swatch{turquoise}~$\ell_1\ell_2$+\MET{}+jet \cite{ATLAS:2024aht}  
  \end{tabular}
  \caption{Scans over the CP-odd higgs mass $\MA$ and (a) \tanb (b) $\kappa$. In both cases,
    \MN = 500 GeV, $V_{mix}=10^{-4}$ and the leptoquarks, CP-even and charged Higgs bosons are decoupled.
    Other details as Fig.~\ref{fig:MHscans1}.
  }
  \label{fig:mascans}
\end{figure}

\subsection{CP-odd Higgs}

The \leff parameter has no effect on the phenomenology of $A$, the CP-odd Higgs. In addition, as with $H$, there
is little dependence on the heavy neutrino mass, $\MN$. Therefore we show, in Fig.~\ref{fig:mascans}, only
the dependence on
$\tanb$ and $\kappa$.
The sensitivity as a function of \tanb is very similar to that for \MH, and again arises from the decays
to $\tau$-leptons,
as expected from Fig.s~\ref{subfig:A1} and \ref{subfig:A3}.

\begin{figure}[hbt]
    \centering
    \includegraphics[width=0.9\textwidth]{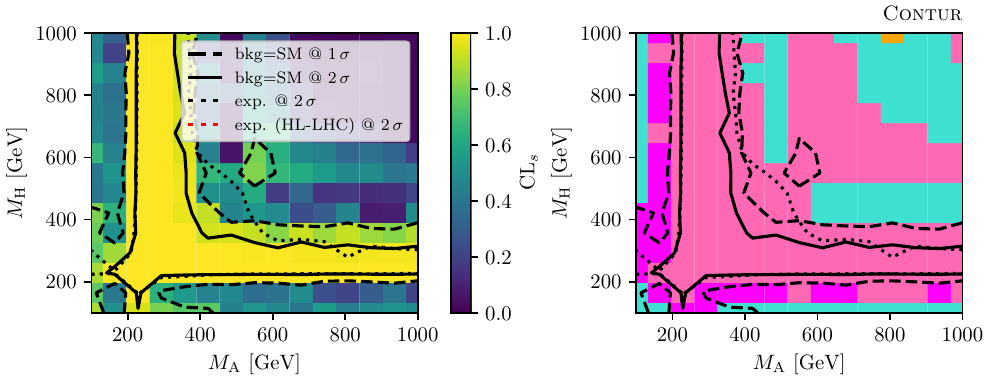}
    \caption{Scan over CP-odd and CP-even Higgs masses, with $\tan\beta=\kappa=\leff=0.1$.
      Right handed neutrino mass, $M_N = 500$ GeV and $V_{mix}=10^{-4}$. The charged Higgs and leptoquarks are decoupled.
      Other details as Fig.~\ref{fig:MHscans1}.
    }
    \begin{tabular}{llll}
      \swatch{magenta}~4$\ell$ \cite{ATLAS:2021kog} & 
      \swatch{hotpink}~$\tau^+\tau^-$ \cite{ATLAS:2025oiy} & 
      \swatch{turquoise}~$\ell_1\ell_2$+\MET{}+jet \cite{ATLAS:2023gsl,ATLAS:2024aht}  &
      \swatch{orange}~$\ell^+\ell^-$+jet
    \end{tabular}
    \label{fig:mamh}
\end{figure}

Before moving on to the charged Higgs boson, we show in Fig.\ref{fig:mamh} an example scan
where both the CP-even and CP-odd Higgses are active, with masses varying independently.
It can be seen that the sensitivity to both is very similar, with the \MA exclusive reaching
slightly higher mass, but both allowed above 400~\GeV. The sensitivity drops away at low masses because
the ATLAS measurement~\cite{ATLAS:2025oiy} has a lower kinematic cutoff,
driven by the need to select high-\pT $\tau$ leptons in the trigger.
We note that the recent search for boosted low-mass states decaying to $\tau$ pairs~\cite{ATLAS:2026txp}
may have an impact in this region.

\subsection{Charged Higgses}

\begin{figure}[hbt]
    \centering
    \includegraphics[width=0.9\textwidth]{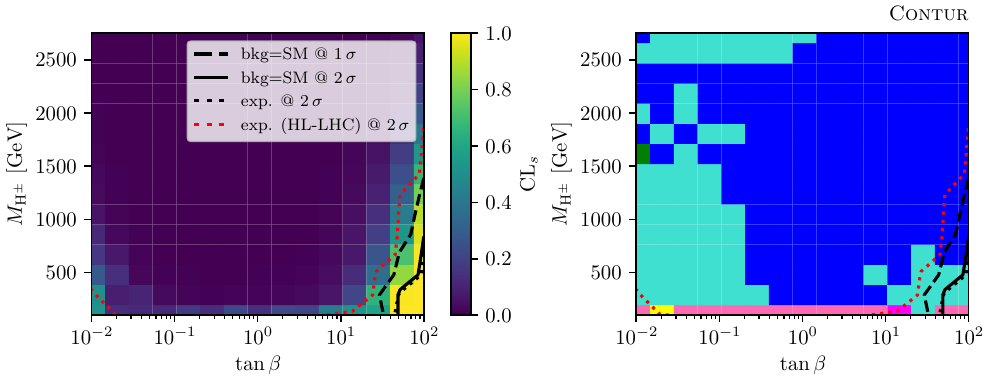}
    \caption{Scan over the charged Higgs mass and $\tanb$.
      Right handed neutrino mass, $M_N = 500$ GeV and $V_{mix}=10^{-4}$. The neutral Higgs bosons and the leptoquarks are decoupled.
      Other details as Fig.~\ref{fig:MHscans1}.
    }
    \begin{tabular}{llll} 
      \swatch{turquoise}~$\ell_1\ell_2$+\MET{}+jet 
      \cite{ATLAS:2024aht} & 
      \swatch{hotpink}~$\tau^+\tau^-$ \cite{ATLAS:2025oiy} & 
      \swatch{blue}~$\ell$+\MET{}+jet \cite{CMS:2021vhb} &  \\ 
      \swatch{green}~\MET{}+jet &
      \swatch{yellow}~$\gamma$  & 
    \end{tabular}
    \label{fig:mhpm_tB}
\end{figure}

Unlike neutral scalars, the charged Higgs bosons cannot be produced by gluon-fusion.
The dominant pair-production processes are electroweak production of the $H^\pm$.
One can also have the associated production with the top quark.
In this case the final state usually contains tau lepton and missing energy, and the di-tau measurement is not sensitive.
The cross section for electroweak pair production of the charged Higgs is generally too small to give sensitivity for charged
Higgs masses high enough for the leptons to pass the fiducial selection of the measurement.
In Fig.~\ref{fig:mhpm_tB}, we show the sensitivity scan for the mass against \tanb, starting the plots at $100$ GeV
to ensure consistency with the LEP2 bounds~\cite{ALEPH:2013htx}. (See also Ref.~\cite{ParticleDataGroup:2026mpi} for more details.)
There is very little sensitivity in current measurements apart from at very high \tanb
where \ttbar measurements provide sensitivity because the cross section increases and the
final state is more often $\ttbar + b$ and $\ttbar+\overline{b}$.
Regarding single production, searches in $\tau$-plus missing energy final states have not been performed in the context
of this model, but typically set cross section limits at a few tens of fb for transverse momenta below about
500~GeV~\cite{CMS:2022ncp,ATLAS:2026iiz} and do not extend to the lower masses considered here.

\section{SUMMARY}
\label{sec:summary}
In this work, we have investigated the Higgs sector phenomenology of the minimal low scale quark--lepton unification framework based on
the gauge symmetry, $SU(4)_C \otimes SU(2)_L \otimes U(1)_R$~\cite{FileviezPerez:2013zmv}.
In addition to the SM-like Higgs boson, this theory predicts a CP-even neutral Higgs, $H$, a CP-odd neutral Higgs, $A$,
and two charged Higgses, $H^\pm$, together with heavy right-handed neutrinos and leptoquarks.
Our analysis focused on the scalar sector of the theory and on the interplay between Higgs decays, heavy neutrino decays,
and collider observables at the LHC.

We discussed in detail the Higgs and Yukawa sectors of the model and emphasized that the couplings of the new Higgs bosons are directly
tied to the structure of matter unification. In the alignment limit, the dominant decays of the heavy Higgs states are typically
controlled by third-generation fermions, 
When kinematically allowed, decays involving heavy right-handed neutrinos can become relevant and may substantially alter the branching
fractions of the extended Higgs sector. We also showed that the SM-like Higgs can have an invisible decay into neutrino states,
providing an additional probe of the model. A notable feature of this framework is that the heavy Higgs decay patterns satisfy
characteristic quark--lepton unification relations in the small- and large-\tanb regimes, offering potentially distinctive
signatures of the underlying theory.

We then studied the collider phenomenology of the new scalar states. We computed the production cross sections and branching ratios as
functions of the Higgs masses and of the relevant effective parameters such as \tanb, $\kappa$, and \leff. The neutral Higgs bosons
can be produced with sizable rates through gluon fusion, while the charged Higgs production is electroweak, and its pair production
cross section is therefore much smaller.
This difference is reflected directly in the collider sensitivity: the current LHC data already place meaningful constraints on the
neutral scalar sector, whereas the reach for the charged Higgs boson is more limited at present.

To quantify the present and future collider sensitivity, we confronted the predicted signals with existing LHC fiducial cross-section
measurements using \herwig, \rivet, and \contur. We found that current measurements already exclude significant regions of parameter
space for the heavy neutral Higgs bosons, especially in regions where decays into third-generation fermions dominate.
In practice, the strongest sensitivity arises from measurements probing $\tau^+\tau^-$, $t\bar t$, and related final states.
The dependence on the heavy neutrino mass is generally modest unless the decay channels into right-handed neutrinos are kinematically
open and competitive. For the charged Higgs boson, current sensitivity remains weak because of the small production cross section,
although the High-Luminosity LHC is expected to improve the reach substantially.
Our results show that Higgs observables provide a powerful and complementary probe of low scale matter unification. While collider
studies of this class of theories have often focused on the leptoquark sector, the extended Higgs sector carries equally important
and in some cases cleaner information about the underlying quark--lepton symmetry.\\

{\small {\textit{Acknowledgments:}}
  This work made use of the High Performance Computing Resource in the Core Facility for Advanced Research
  Computing at Case Western Reserve University and the HEP compute cluster at UCL.
\newpage
\appendix
\section{APPENDIX}
\label{appendix}
The relevant Feynman rules  are given by
\begin{eqnarray}
    \bar{u}_iu_jH & : &  \hspace{0.0 cm} i\  \left[(C_{uu}^H )_{ij} P_R + (C_{uu}^H)^*_{ji} P_L\right], \\
    \bar{d}_i d_j H & : &  \hspace{0.0 cm} i\  \left[(C_{dd}^H )_{ij} P_R + (C_{dd}^H)^*_{ji} P_L\right], \\
    \bar{e}_ie_j H & : & \hspace{0.0 cm} i\  \left[(C_{ee}^H )_{ij} P_R + (C_{ee}^H)^*_{ji} P_L\right], \\
    \bar{\nu}_iN H & : & \hspace{0.0 cm} i \left[(C_{N\nu }^H)_{ij} P_R +(C_{N\nu}^H)^*_{ji} P_L\right], \\
     \bar{u}_iu_jA & : & \left[(C_{uu}^A)_{ij} P_R - (C_{uu}^A)_{ji}^* P_L \right], \\
    \bar{d}_id_jA & : & \left[(C_{dd}^A)_{ij} P_R - (C_{dd}^A)_{ji}^* P_L \right], \\
    \bar{e}_ie_jA & : & \left[(C_{ee}^A)_{ij} P_R - (C_{ee}^A)_{ji}^* P_L \right],\\
    \bar{\nu}_iN_j A & : & \left[(C_{N\nu}^A)_{ij} P_R - (C_{N\nu}^A)^*_{ji} P_L  \right], \\
    \bar{u}_id_jH^+ & : & i\hspace{0.0 cm} \left[(C_{ud}^{L})_{ij} P_L + (C_{ud}^R)_{ij} P_R \right],\\
    \bar{\nu}_i e H^+ & : & -i C_{\nu e }^{H^+} P_R, \\
     H^+\bar{N}\ e & :  & i\  C_{Ne}^{H^+} P_L , \\
     hhH & : & i \leff \ v,
\end{eqnarray}
where the coefficients are 
\begin{eqnarray}
    C_{uu}^H & = & U_L^\dagger \left[\frac{Y_1}{\sqrt{2}} \cos\alpha + \frac{Y_2}{2 \sqrt{6}} \sin{\alpha}\right] U_R, \\
    C_{uu}^A & = & U_L^\dagger \left[ \frac{Y_1}{\sqrt{2}}\sin{\beta} - \frac{Y_2}{2\sqrt{6}}\cos\beta\right]U_R, \\
    C^H_{dd} & = & D_L^\dagger \left[\frac{Y_3}{\sqrt{2}}\cos{\alpha}+ \frac{Y_4}{2\sqrt{6}}\sin{\alpha}\right]D_R, \\
    C_{dd}^A & = & D_L^\dagger \left[ \frac{Y_3}{\sqrt{2}}\sin{\beta} - \frac{Y_4}{2\sqrt{6}}\cos\beta\right]D_R, \\
    C_{ee}^H & = & E_L^\dagger \left[\frac{Y_3}{\sqrt{2}} \cos{\alpha} - \frac{3}{2 \sqrt{6}}Y_4\sin{\alpha}\right]E_R, \\
    C_{ee}^A & = & E_L^\dagger \left[ \frac{Y_3}{\sqrt{2}}\sin{\beta} + \frac{3}{2\sqrt{6}}Y_4\cos\beta\right]E_R, \\
    C_{N\nu}^{H} & = & N_L^\dagger \left[\frac{Y_1}{\sqrt{2}} \cos{\alpha} - \frac{3}{2 \sqrt{6}}Y_2 \sin{\alpha}\right]N_R, \\ 
    C_{N\nu}^A & = & N_L^\dagger \left[ \frac{Y_1}{\sqrt{2}}\sin{\beta} + \frac{3}{2\sqrt{6}}Y_2\cos\beta\right]N_R, \\
    C_{ud}^L & = & U_R^\dagger \left[Y_1^\dagger \sin\beta - \frac{Y_2^\dagger}{2\sqrt{3}}\cos\beta\right]D_L, \\
    C^R_{ud} &= & U_L^\dagger\left[-\sin{\beta}\  Y_3 + \frac{Y_4}{2\sqrt{3}} \cos\beta\right]D_R, 
    \end{eqnarray}
    \begin{eqnarray}
    C_{\nu e}^{H^+} &= & N_L^\dagger \left[Y_3 \sin{\beta} + \frac{3}{2\sqrt{3}} Y_4 \cos{\beta} \right] E_R, \\
    C_{N e}^{H^+} & = & N_R^\dagger \left[Y_1^\dagger \sin{\beta} + \frac{\sqrt{3}}{2}Y_2^\dagger \cos{\beta} \right]E_L.
\end{eqnarray}
In the alignment limit, $\sin{(\beta -\alpha)} \to 1$ , the coefficients are given by 
\begin{eqnarray}
    C_{dd}^H & = & (3 \tanb - \cot\beta) \frac{M_D^{diag}}{4 v} + (\tanb + \cot \beta ) \frac{V_L M_E^{diag} V_R^\dagger}{4 v} \label{fveqn2}, \\
    C^H_{ee} &= & (\tanb -3\cot{\beta})\frac{M_E^{diag}}{4v}+ 3(\tanb+\cot{\beta})\frac{V_L^\dagger  M_D^{diag}V_R}{4v}, \\
     C_{dd}^A & = & (3 \tanb - \cot\beta) \frac{M_D^{diag}}{4 v} + (\tanb + \cot \beta ) \frac{V_L M_E^{diag} V_R^\dagger}{4 v},\\
      C^A_{ee} &= & (\tanb -3\cot{\beta})\frac{M_E^{diag}}{4v}+ 3(\tanb+\cot{\beta})\frac{V_L^\dagger  M_D^{diag}V_R}{4v},\\
      C_{\nu e}^{H^+} &= & 3(\tanb+\cot{\beta})\frac{V_2^\dagger  M_D^{diag}V_3}{2 \sqrt{2}v}+ (\tanb-3\cot{\beta})\frac{V_4  M_E^{diag}}{2\sqrt{2}v},
      \label{fveqn}
\end{eqnarray}
where the mixing matrices are defined as 
\begin{eqnarray}
    V_L = D_L^\dagger E_L , \hspace{0.5cm} V_R = D_R^\dagger E_R , \hspace{0.5 cm} V_2 = D_L^\dagger N_L , \hspace{0.5 cm} V_3 = D_R^\dagger E_R, \hspace{0.5cm} V_4 = N_L^\dagger E_L.
\end{eqnarray}
Here, for simplicity, we assume the matrices $V_L, V_R,V_2,V_3,V_4 \sim \mathbf{1}$. For the constraints from flavour violating processes, see the study in Ref.~\cite{FileviezPerez:2022fni}.
Given the flexibility in choosing the Yukawa couplings $Y_1$, $Y_2$ and the mixing matrices, we parametrize the coupling of up quarks with a free parameter $\kappa$ as follows 
\begin{eqnarray}
    C_{uu}^{H,A} =C^{H,A}_{N\nu}= \frac{1}{\sqrt{2}}C_{N e}^{H^+}= \frac{1}{\sqrt{2}} C_{ud}^L = \frac{\kappa}{4 v} M_U^{\text{diag}} \hspace{0.2 cm} \text{and}   \hspace{0.2 cm} \frac{1}{\sqrt{2}}C_{ud}^R = \kappa \ C_{dd}^H. 
\end{eqnarray}
In the decoupling limit, the masses for the Higgses can be written as 
\begin{eqnarray}
    \MH^2 & = &\frac{m_{12}^2}{s_\beta c_\beta} +   \  v^2  \left[\lambda_1 s_\beta^2 c_\beta^2 + \lambda_2 s_\beta^2 c_\beta^2 - 2(\lambda_3+\lambda_4+\lambda_5)s_\beta^2 c_\beta^2) -\lambda_6(\frac{1}{2} ct_\beta + 2 s_\beta c_\beta c_{2\beta}) 
    -\lambda_7 (\frac{1}{2} t_\beta - 2 s_\beta c_\beta c_{2\beta})\right], \nonumber \\ \\
    \Mh^2  &=&   \ v^2 (\lambda_1 c_\beta^4 + \lambda_2 s_\beta^4 + 2(\lambda_3+\lambda_4+\lambda_5)s_\beta^2 c_\beta^2 + 2 \lambda_6 c_\beta^2 s_{2\beta} + 2 \lambda_7 s_\beta^2 s_{2 \beta}),\\
   \MA^2 &=& \frac{m_{12}^2}{s_\beta c_\beta} - \frac{v^2}{2} (2\lambda_5 + \lambda_6 ct_\beta + \lambda_7 t_\beta),\\
   M_{H^{\pm}}^2 &=& \MA^2 + \frac{v^2}{2} (\lambda_5 -\lambda_4  ),
\end{eqnarray}
where $\tanb = t_\beta = v_2/v_1$,  $c\beta= \cos{\beta} , \  s_\beta = \sin{\beta} ,\ \text{and} \ ct_\beta = \cot{\beta}$. 
For simplicity, we define the parameter $\lambda_H $, $\lambda_h$ and $\lambda_A$ as 
\begin{eqnarray}
    \lambda_H &=& \lambda_1 s_\beta^2 c_\beta^2 + \lambda_2 s_\beta^2 c_\beta^2 - 2(\lambda_3+\lambda_4+\lambda_5)s_\beta^2 c_\beta^2) -\lambda_6(\frac{1}{2} ct_\beta + 2 s_\beta c_\beta c_{2\beta}) 
      -\lambda_7 (\frac{1}{2} t_\beta - 2 s_\beta c_\beta c_{2\beta}), \nonumber \\ 
      \end{eqnarray}
      \begin{eqnarray}
     \lambda_h  &=& (\lambda_1 c_\beta^4 + \lambda_2 s_\beta^4 + 2(\lambda_3+\lambda_4+\lambda_5)s_\beta^2 c_\beta^2 + 2 \lambda_6 c_\beta^2 s_{2\beta} + 2 \lambda_7 s_\beta^2 s_{2 \beta}), \\
     \lambda_A &=& \frac{1}{2}(2\lambda_5 + \lambda_6 ct_\beta + \lambda_7 t_\beta).
\end{eqnarray}

\bibliography{ref}
\end{document}